\documentclass[%
 reprint,
superscriptaddress,
 amsmath,amssymb,floatfix,
]{revtex4-2}
\usepackage{graphicx}
\usepackage{dcolumn}
\usepackage{comment}
\usepackage{bm}
\usepackage{color}
\usepackage{lipsum}

\DeclareMathOperator\erf{erf}

\begin{document}
\title{Enhanced Coalescence in Driven Foams}

\author{Alice Requier}
\affiliation{Universit\'e Paris-Saclay, CNRS, Laboratoire de Physique des Solides, 91405 Orsay, France}
\author{Andrea Plati}
\affiliation{Universit\'e Paris-Saclay, CNRS, Laboratoire de Physique des Solides, 91405 Orsay, France}
\author{Emmanuelle Rio}
\affiliation{Universit\'e Paris-Saclay, CNRS, Laboratoire de Physique des Solides, 91405 Orsay, France}
\author{Anniina Salonen}
\affiliation{Universit\'e Paris-Saclay, CNRS, Laboratoire de Physique des Solides, 91405 Orsay, France}
\affiliation{Soft Matter Sciences and Engineering, ESPCI Paris, PSL University, CNRS, Sorbonne Université, 75005 Paris, France}
\date{\today}

\begin{abstract}

External driving leads to the emergence of unique phenomena and properties in soft matter systems. We show that driving quasi-2D foams by mechanical vibration results in significant bubble coalescence, which is enhanced by the continuous phase yield stress. The competition between coarsening and coalescence can be modulated through vibration amplitude and foam liquid fraction, which can be used to create unusual structural motifs. The combined effect of coarsening and coalescence is captured through a statistical model that quantitatively describes the time evolution of the number of bubbles.
\end{abstract}

\maketitle

\emph{Introduction---}
Soft matter systems exhibit unique phenomena when subjected to external driving \cite{nagel2017experimental}. This is the case of shear thickening suspensions \cite{seto2013discontinuous,fall2015macroscopic}, vibrofluidised granular materials \cite{umbanhowar1996localized,dijksman2011jamming,plati2024quasi} or active colloids \cite{hayward2000electrophoretic,wurger2008transport,bricard2013emergence,wei2023reconfiguration}. The macroscopic properties of these  systems depend significantly on the combined effects of constituting materials and driving mechanisms. All this becomes even more complex when considering the coexistence of multiple phases, a condition that is nevertheless ubiquitous in the natural environment and applications.

Liquid foams are a paradigmatic case of multiphase matter composed of gas bubbles dispersed in a continuous fluid phase~\cite{cantat2013foams}. They are unstable in time as their average bubble size grows through both coarsening and coalescence processes. A foam coarsens as gas transfers between neighbouring bubbles because of differences in Laplace pressure, eventually leading to a self-similar growth regime and a homogeneous structure~\cite{Mullins86, vonNeumann_Coarsen, Kraynik2001}. As for coalescence, it is caused by the rupture of thin films between bubbles and can result in catastrophic foam collapse and highly heterogeneous foam morphologies~\cite{Langevin2025}. As both phenomena influence foam structure and are influenced by it, it is difficult to disentangle their contributions to foam ageing.

During many processes, foams undergo mechanical perturbations, such as shear or vibrations, which may alter their ageing process. If the foam is sheared enough to promote significant rearrangements, the average film thickness increases and coarsening slows considerably \cite{saint2023foam}. However, vigorous shearing can promote bubble coalescence in 2D foams  \cite{Mohammadigoushki_shear_coalesce2D}. As for mechanical or ultrasonic vibrations, they also lead to foam destabilisation by promoting coalescence or accelerating drainage~\cite{Morey1990,Sandor1993}, making them good candidates in the design of foam breakers~\cite{Barigou2001}. The enhancement of coalescence induced by vibrations could be due to the local deformation of the liquid channels \cite{Elias2020}.  

In order to improve foam stability, yield stress fluids are commonly used in foods and cosmetics, as they can hinder bubble rearrangements~\cite{Guidolin2024}, leading to slower coarsening rates~\cite{Requier2024,Galvani2025}, and can eventually result in unusual bubble patterns~\cite{guidolin2023viscoelastic}. A foam continuous phase with a sufficiently high yield stress can even arrest bubble growth~\cite{Lesov2014,Galvani2025}.

In this Letter, we investigate the interplay between coarsening and coalescence by mechanically vibrating a foamy yield stress fluid. We show in particular that the destabilising impact of vibrations is exacerbated by the presence of the yield stress fluid. The driving amplitude can be used to modulate the relative rates of coarsening and coalescence, which can be described using the statistical model that we propose. Combined tuning of the vibration amplitude and the foam liquid fraction gives us control of the resulting foam morphology.

\begin{figure*}
\centering
\includegraphics[width=0.99\textwidth,clip=true]{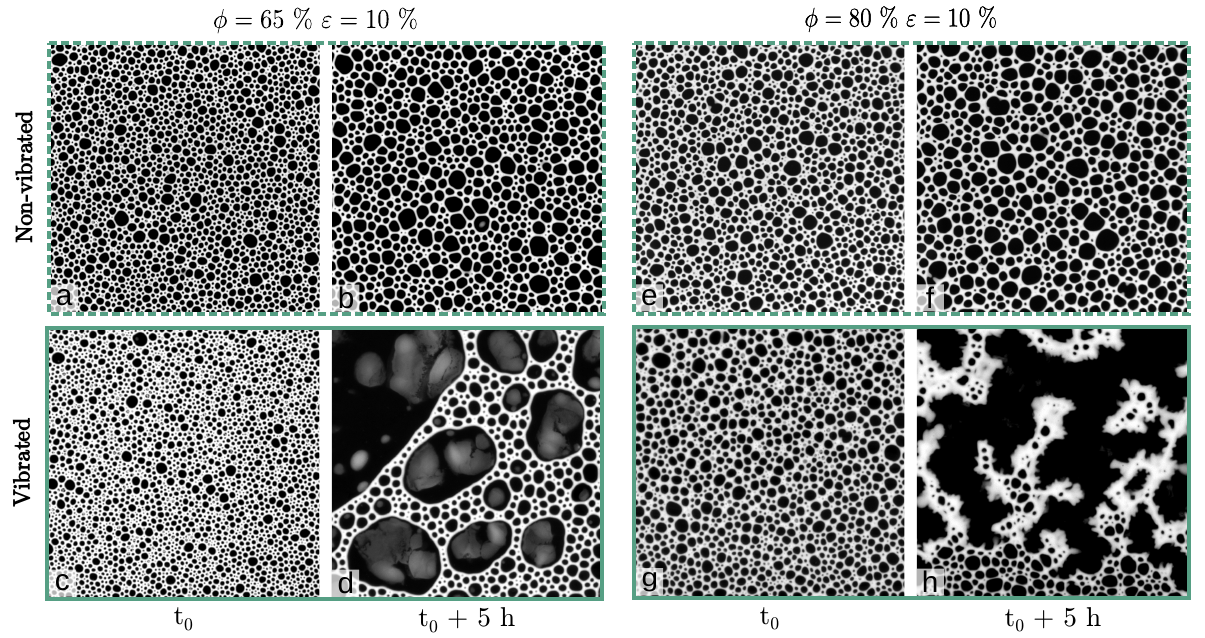}
\caption{Snapshots of the system's structure once the foam becomes 2D (time $t_0$) and after 5 hours for vibrated ($f$ = 53 Hz, $A$ = 0.26 mm) and non-vibrated foamed emulsions with $\phi=\{65,80\}$ $\%$. The emulsion with oil fraction $\phi=80$ $\%$ has a yield stress of $\tau_c\sim 16.5$ Pa, while none is measured at $\phi=65$ $\%$. The edge of each photograph is 70 mm.}
\label{fig:Fig1}
\end{figure*}

\emph{Experimental setup---} We work with quasi-2D foams consisting of a single layer of bubbles confined between two horizontal plates 1 mm apart. 
This configuration eliminates gravitational drainage and enables direct visualisation of the foam structure during destabilisation. The continuous phase is an emulsion composed of an aqueous solution of Sodium Dodecyl Sulfate at 30 g/L and drops of sunflower oil (both from Sigma Aldrich), which are produced with the double-syringe technique~\cite{gaillard2017controlled}. Depending on the oil volume fraction in the emulsion $\phi$, it can behave as a viscous fluid ($\phi = 65\%$) or exhibit a yield stress ($\phi = 80\%$, $\tau_c\sim 16.5$~ Pa \cite{alicethese}). Air is incorporated into the emulsions using a planetary motion food processor (Kenwood KMIX750AW, 1000 W) to make foams at varying liquid fractions $\varepsilon\in [8,20]$ $\%$. 

Once ready, the foam is sealed between two plates attached to an electrodynamic shaker (Brüel
\& Kjær, LDS V400). A sinusoidal electric signal is first sent to an amplifier (Brüel \& Kjær, LDS LPA600) and then to the shaker, which imposes a vertical displacement $z_{\rm{p}}(t)=A\sin(2\pi f t)$ to the cell. In all our experiments we kept $f=53$ Hz and varied the vibration amplitude $A\in [0,0.26]$ mm, which gives a range of peak acceleration of $\Gamma=(2\pi f)^2A/g\in [0.1, 2]$. 
At the beginning of the experiment, vibrations are switched on, and the bubbles start to grow. After a time $t_0$ they become big enough to form a single layer and the foam is considered as 2D \cite{guidolin2023viscoelastic}. In the following, we plot everything as a function of $t-t_0$.
Snapshots of the foam structure have been acquired every 30 seconds with a high-resolution camera (Basler a2A2590 - 60 $\mu$m, 3840 × 2748 pixels$^2$) equipped with a lens (Fujinon 12 mm 1:1.8) and positioned above the setup. Image processing is performed using custom MATLAB scripts. First, foam images are segmented using adaptive thresholding, and then their structures are skeletonised through a watershed algorithm. This allows us to measure the radius of each bubble and their total number. 

\emph{Foam destabilisation---} 
In Fig.~\ref{fig:Fig1}, we compare the evolution of foamed emulsions at $\phi=65$~$\%$ and $\phi=80$ $\%$ with and without applying mechanical vibrations, at a fixed liquid fraction $\varepsilon=10$ $\%$. We show how the coarsening evolution affects the foam structure over time by comparing snapshots taken at the initial time $t_0$ and after 5 hours. In the studied time frame, non-vibrated foams at $\phi=\{65,80\}$ $\%$ showed qualitatively similar results, where the structure evolves mainly through coarsening (Figs.~\ref{fig:Fig1}a, b, e and f).  

In panels c and d, we show the comparison for a vibrated foam with $\phi=65$ $\%$. In this case, we observed an initial evolution mainly characterised by coarsening, followed by a regime with coalescence events homogeneously distributed in time (see video in the Supplemental Materials). From panel d, we also note that coalescence leads to a more advanced state of destabilisation at time $t_0+5$~h characterised by a few very large bubbles surrounded by foam. However, locally, we find an equilibrium topology \textit{i.e.} Plateau's laws are respected~\cite{vonNeumann_Coarsen}. 

When considering the vibrated foam with $\phi=80$ $\%$ (Figs.~\ref{fig:Fig1}g and h), the scenario changes qualitatively. In this case, we observe an initial coarsening regime followed by a cascade of coalescence events, which dramatically accelerate foam destabilisation (see video in the Supplemental Materials). Moreover, after five hours, the surface covered by bubbles is drastically reduced and most of the liquid phase is distributed in heterogeneous branched structures.  

It is now important to note that 65 $\%$ and 80 $\%$ oil fraction emulsions differ in their rheological properties. The former behaves as a purely viscous fluid without exhibiting any yield stress in rheological tests, the latter is instead characterised by a yield stress $\tau_c\sim 16.5$~ Pa, meaning that when subjected to an external stress $\tau < \tau_c$ it responds elastically to the resulting deformation. Our experimental results suggest that the coalescence catastrophe is triggered by the combined effect of mechanical vibration and yield stress in the continuous phase. To further support this picture, we also performed some experiments with aqueous foams (see End Matter) and verified that they experience no coalescence catastrophe. 

In other words, regardless of the specific value of $\phi$, foamed emulsions behave similarly to aqueous foams while subjected to mechanical vibration as long as there is no yield stress.  
In this letter, we are particularly interested in the observed enhanced coalescence dynamics, so we will mainly focus on foamed emulsions at $\phi=80$ $\%$.

\begin{figure}
\centering
\includegraphics[width=0.98\columnwidth,clip=true]{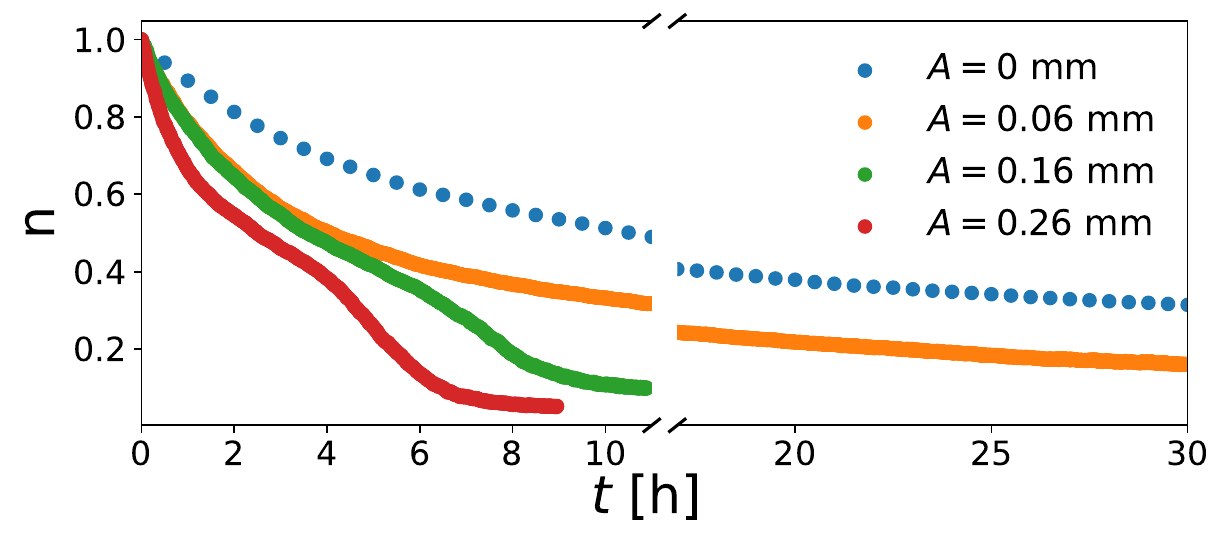}
\caption{Rescaled number of bubbles as a function of time for different shaking amplitudes in foamed emulsion with $\varepsilon=10$ $\%$ and $\phi=80$ $\%$. }
\label{fig:Fig2}
\end{figure}

We found that a simple observable able to discriminate the different effects of coarsening and coalescence is the time evolution of the rescaled number of bubbles $n(t)=N(t)/N(0)$, where $N$ is the instantaneous number of bubbles in the foam. In Fig.~\ref{fig:Fig2}, we show $n(t)$ for different vibration amplitudes with $\varepsilon=10$ $\%$ and $\phi=80$~$\%$. For $A=0$ and $A=0.06$~mm, the foams mainly evolve by coarsening, with rare coalescence events and $n(t)$ follows a simple decay with positive convexity. 
The decay rate is found to be larger for the larger amplitude. For $A=0.16$~mm and $A=0.26$~mm, we observe instead that after an initial regime similar to the one observed for lower vibrations, the curve develops an inflection point followed by a decay with negative convexity in correspondence with the time where the coalescence cascade is observed in the experiments. The last part of the decay is again with positive convexity, and this corresponds to the final evolution of the foam when the coalescence cascade has stopped and the liquid phase is distributed in the branched structure (see Fig.~\ref{fig:Fig1}h). The behaviour of $n(t)$ for different vibration amplitudes tells us that pure coarsening is a process which slows down (decay with positive convexity) and the coalescence cascade is instead self-amplified (decay with negative convexity). The long time behaviour for $A=0$~mm and $ A=0.06$~mm is provided to show that the inflection point does not occur at later times.
Another important observation is that, for all $A$, the initial ageing mechanism is always pure coarsening and the coalescence cascade is triggered at a given average bubble size. Comparing the curves for $A=0.16$~mm and $A=0.26$ mm, we see that the onset of the cascade is anticipated for larger vibration amplitudes. To summarise, Fig.~\ref{fig:Fig2} shows that foam destabilisation can be controlled by mechanical vibrations in two ways; for weak vibrations one can speed up coarsening without causing coalescence cascade, for larger amplitudes one can trigger the coalescence cascade and tune the time of its start. To complement this analysis based on $n(t)$, in the End Matter we also show the time evolution of the average bubble radius.

\begin{figure}
\centering
\includegraphics[width=0.99\columnwidth,clip=true]{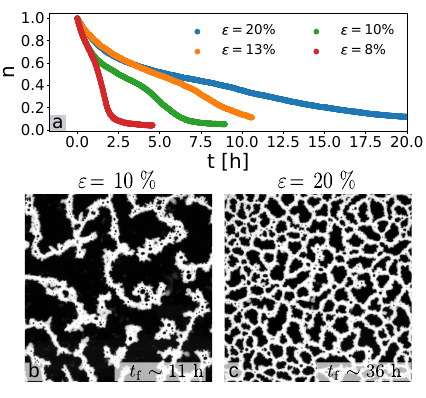}

\caption{a) Rescaled number of bubbles as a function of time for different liquid fractions in foamed emulsion with $\phi=80$ $\%$ vibrated at amplitude $A=0.26$ mm. System's structure at the final time $t_{\rm f}$ of the experiment for b) $\varepsilon=10$ $\%$ ($t_{\rm f}\sim 11$ h) and c) $\varepsilon=20$ $\%$ ($t_{\rm f}\sim 36$ h).}
\label{fig:Fig3}
\end{figure}

We now turn our attention to the effect of liquid fraction on the vibrated foams. In Fig.~\ref{fig:Fig3}a, we plot $n(t)$ for different values of $\varepsilon$ with $\phi=80$ $\%$ and $A=0.26$ mm. We note that increasing $\varepsilon$ delays the inflection point in the curve. By visual inspection of foam evolution we observed that this corresponds to a less abrupt coalescence cascade (see video in the Supplemental Materials). Note that in all these experiments, after the initial coarsening regime, the main observed destabilisation mechanism is always coalescence. This means that for large vibration amplitudes ($A=0.26$ mm) it is not possible to restore pure coarsening by playing with the liquid fraction but only to make the coalescence cascade less abrupt. A particularly interesting effect of liquid fraction we observed regards the spatial organisation of the liquid phase at the end of the experiments (\textit{i.e.} when the number of bubbles ceases to evolve in time). In Figs.~\ref{fig:Fig3}b and c, we show the final snapshots of a foamed emulsion with respectively 10 \% and 20 \% liquid fraction. In the first case, we see the final state of the foam from Fig.~\ref{fig:Fig1}h where the bubbles have completely disappeared and all the fluid phase forms relatively isolated branches. For $\varepsilon=20$ $\%$, we observe that there is enough liquid in the system to allow different final branches to merge and form a new ``percolating foam structure" which spans all over the system and presents a characteristic topology radically different with respect to a foam evolved by pure coarsening. 

We point out that the coincidence of ageing arrest with these characteristic structures appears to be related to the inability of shaking to redistribute liquid when it forms thick blocks, as can be seen in Figs.~\ref{fig:Fig3}b and 3c. Considering that the liquid phase must undergo a minimum stress $\tau_c$ to yield and that the stress experienced by the continuous phase between the two parallel plates moving in phase depends on its deformation, we can qualitatively explain the observed behaviour: thicker liquid blocks are stiffer and thus deform less when shaken at a given amplitude.

\emph{A model for foam destabilisation---}
In this section, we present a  model for the temporal evolution of $n$, which accounts for the competition between coarsening and coalescence. The proposed model is as follows: 
\begin{equation}
    \dot{n}=-ap(n)n-bn^2.\label{eq::modII}
\end{equation}
The first term accounts for coalescence while the second one describes standard coarsening at rate $b$ \cite{Mullins86}. To model the coalescence term, we take inspiration from a previous study that focused on the stochastic nature of coalescence events in foams \cite{Forel2019}. In particular, it was observed that the probability of a contact film between two bubbles breaking increases with the film area. Assuming that the film area scales linearly with the bubble diameter $d$, we express the probability of rupture as $\Tilde{p}(d)$. Then we assume a mean field-like regime where all bubbles have a diameter equal to the mean one $\bar{d}=\sqrt{A_{\rm{tot}}/N}=\alpha/\sqrt{n}$, where $A_{\rm{tot}}$ is the fixed total area of the system and $\alpha=\sqrt{A_{\rm{tot}}/N(0)}$. The probability of a film breaking as a function of the rescaled number of bubbles is then $p(n)=\Tilde{p}(\alpha/\sqrt{n})$. In an infinitesimal time interval $\text{d}t$, the number of bubbles that break is $ap(n)N\text{d}t$, where the parameter $a$ is the inverse of a typical time related to coalescence. So if only coalescence is involved, foam destabilisation would be governed by $\dot{n}=-ap(n)n$. To propose a functional form for $p(n)$, we consider the idea of a resistance function $f_{\rm R}(n)$ which monotonically increases in $n$. Indeed, for a fixed total area, the more bubbles there are, the smaller they are and small bubbles are less likely to break up~\cite{Forel2019}.
With such a resistance function, we define $p(n)$ as the probability that the absolute value of a random variable drawn from a Gaussian distribution with standard deviation $\sigma$ is greater than $f_{\rm R}(n)$. So $p(n)=\text{Prob}(|r_\sigma|>f_{\rm R}(n))$. This probability is associated with the Gaussian cumulative distribution function, which reads:

\begin{equation}
p(n)=2\int_{-\infty}^{-f_{\text{R}}(n)}P_\sigma(r)\text{d}r=1+\erf \left[ \frac{-f_{\text{R}}(n)}{\sqrt{2}\sigma} \right], \label{eq::pn} 
\end{equation} 

where $P_\sigma(r)$ is the normalised Gaussian and $\erf$ is the error function. 
The explicit form of $f_{\rm R}(n)$ is not known and should be proposed based on empirical evidence. With the employed setup, we are not able to access this kind of information, so we propose a simple form $f_{\rm R}(n)\propto n^\lambda$ (\textit{i.e.} $f_{\rm R}(d)\propto d^{-2\lambda}$) with $\lambda=1$. We chose this particular value for $\lambda$ because, as we will see, it allows us to systematically fit all our experimental datasets. We again stress that we do not have a physical justification for this, as the film-breaking mechanism is a rather complex phenomenon that has yet to be fully understood~\cite{langevin2019coalescence}. Regarding the coarsening term of Eq.~\eqref{eq::modII}, it is modeled following a standard mean field theory for dry foams~\cite{Mullins86}. 
The proposed model is then given by Eq.~\eqref{eq::modII} with $p(n)$ from Eq.~\eqref{eq::pn} and depends on three parameters: $a$, $b$, and $\sigma$. To fit them, we differentiated the experimental data to obtain $\dot{n}(t)$ and then $\dot{n}(n)$, which can be directly fitted using Eq.~\eqref{eq::modII}. 
More details on the fitting procedure are given in the End Matter; here we only mention that $b$ is independently determined from the early decay of $n(t)$, which is dominated by coarsening across all experiments. With $b$ fixed, $a$ and $\sigma$ are fitted from $\dot{n}(n)$.

\begin{figure}
\centering
\includegraphics[width=0.99\columnwidth,clip=true]{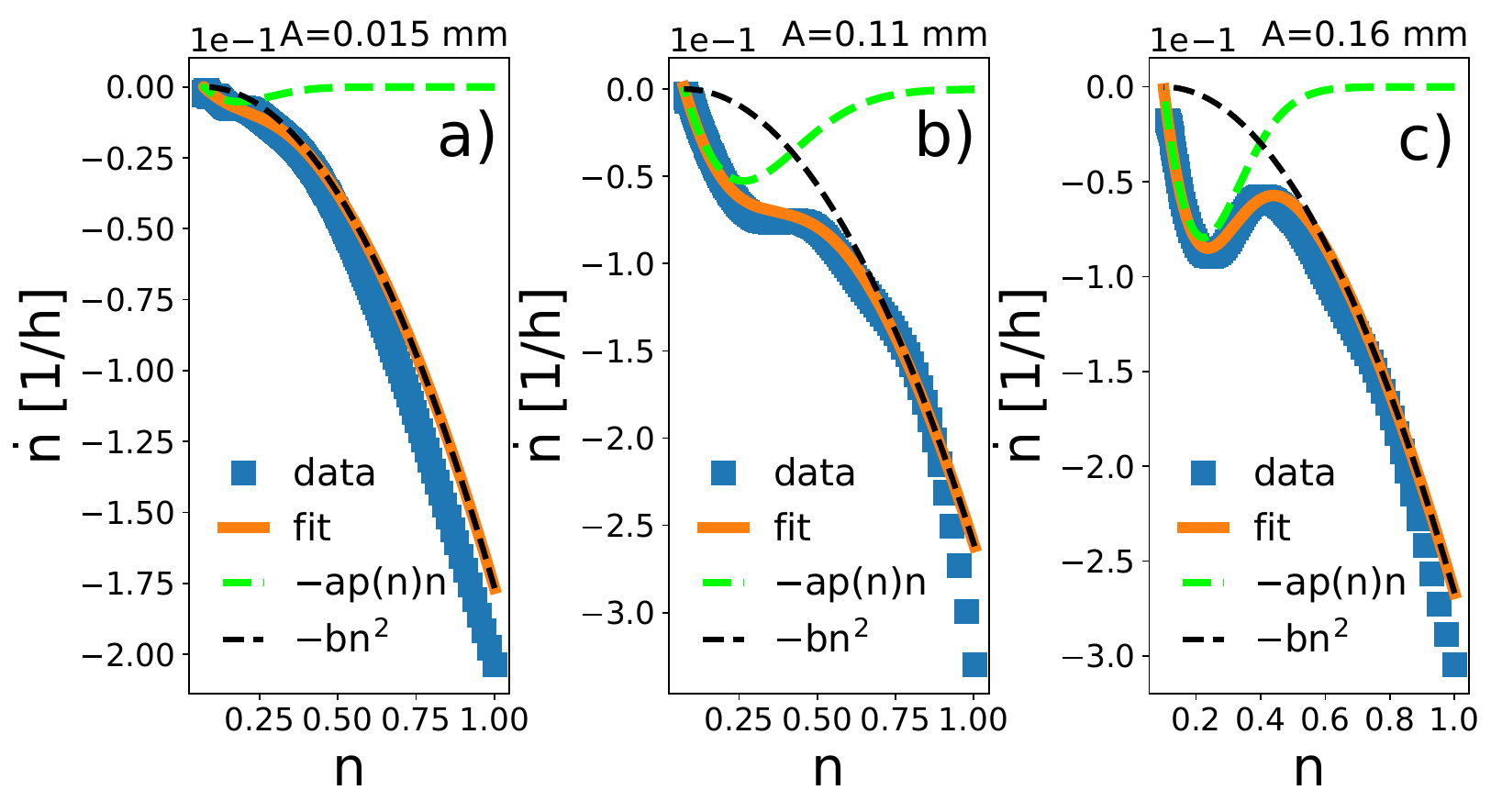}
\includegraphics[width=0.99\columnwidth,clip=true]{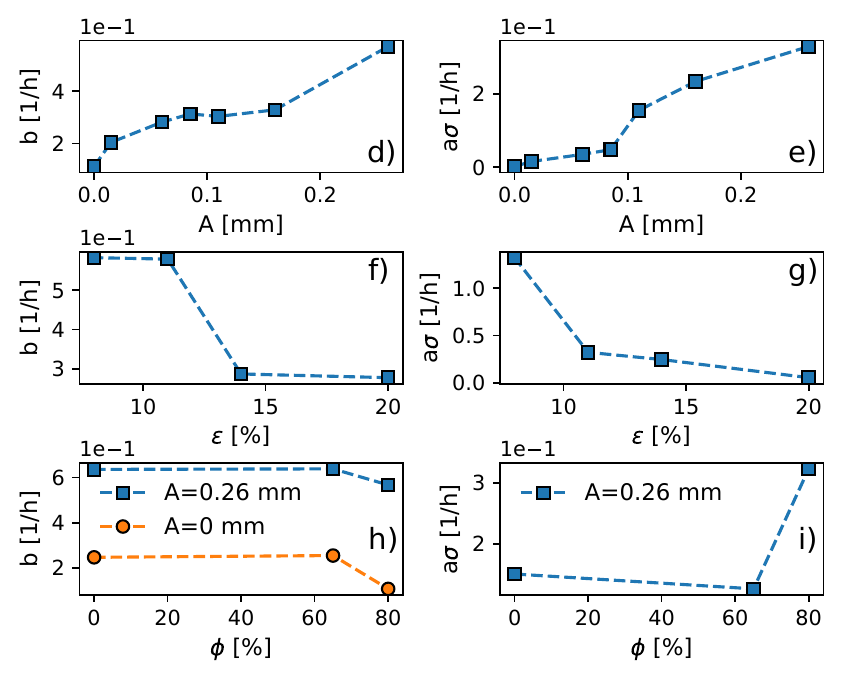}

\caption{Panels a, b and c: derivative of the rescaled number of bubbles as a function of the number of bubbles for three different shaking amplitudes, $\varepsilon=10\%$ and $\phi=80\%$. The data $\dot{n}(t)$ (blue squares) are obtained by differentiating a high-order polynomial fit of the $n(t)$ curves. We provide a comparison between experimental data and the model prediction obtained by fitting $\dot{n}(t)$ with Eq.~\eqref{eq::modII}. We also highlight the resulting contribution of coarsening and coalescence in the model. Panels d, e, f, g, h and i: coarsening rate $b$ and the product $a\sigma$ obtained by fitting Eq.~\eqref{eq::modII} over different experimental curves obtained varying $A$, $\varepsilon$ and $\phi$. Fixed control parameters are $\varepsilon=10\%$  in panels d, e, h, i; $\phi=80$ $\%$ in d, e, f, g; $A=0.26$ in f, g and i.}
\label{fig:Fig4}
\end{figure}

The top panels of Fig.~\ref{fig:Fig4} show representative fits of $\dot{n}(n)$ for three shaking amplitudes. At low amplitude ($A=0.06$ mm, orange curve in Fig.~\ref{fig:Fig2}), the trend is a monotonic, parabola-like curve, consistent with coarsening-driven destabilisation; coalescence events are indeed extremely rare for this shaking amplitude. By increasing $A$, an inflection point appears at $A=0.11$ mm and a non-monotonic shape with a local minimum and maximum is found at $A=0.16$ mm. Notably, our model captures all these features semi-quantitatively.
By plotting the separate contributions of coarsening ($-bn^2$) and coalescence ($-ap(n)n$), we reveal their interplay. Initially ($n \sim 1$), foam destabilisation is driven solely by coarsening, as small bubbles resist coalescence. As bubbles grow and $n$ decreases, coalescence becomes significant, leading to a peak in its contribution (Figs.~\ref{fig:Fig4}b and c) \textit{i.e.} the coalescence catastrophe observed experimentally. We note that this behaviour requires a non-linear $p(n)$: it must be negligible at $n \sim 1$ and rise sharply below a threshold $n^\star$, corresponding to a characteristic bubble size.

To assess the roles of coarsening and coalescence across all control parameters, we systematically fitted our complete datasets obtained with different $A$, $\varepsilon$, and $\phi$. The effect of coarsening is quantified by $b$. It increases as a function of $A$ (panel d) and decreases as a function of $\varepsilon$ (panel f) for $\phi=80\%$ . In panel h, we show that the behaviour of $b$ is unaffected by $\phi$ until the emulsion develops yield stress ($\phi=80$ $\%$). At this point, $b$ drops to lower values for both vibrated and non-vibrated foams.  
Coalescence contribution is captured by the combined parameter $a\sigma$. 
Its behaviour reflects our two main experimental observations. For a yield stress foamed emulsion ($\phi=80$ $\%$), coalescence is strengthened for increasing $A$ (panel e) and weakened for increasing $\varepsilon$ (panel g). However, if the emulsion lacks yield stress $(\phi< 80$ $\%)$, coalescence becomes negligible (panel i). In summary, the effect of control parameters on foam ageing depends on the specific evolution mechanism examined. Vibration amplitude and liquid fraction have the same effect on coarsening and coalescence: the former speeds up ageing and the latter slows it down. The oil fraction, however, acts differently on coarsening and coalescence. Below the yield stress threshold, coarsening is enhanced and coalescence is hindered. As yield stress develops, the opposite occurs.   

\emph{Conclusion---} 
In conclusion, we have shown that mechanical vibrations accelerate foam ageing, in particular coalescence. When foams are made from yield stress fluids, which usually stabilise foams, the impact of vibrations is enhanced. This allows us to modulate the relative rates of coarsening and coalescence and to trigger coalescence cascades. 
We quantify this competition by combining classical coarsening with a statistical model for the coalescence rate. 
The peculiar ageing dynamics we obtain is also linked to the emergence of unusual foam structures. Our study demonstrates the power of external driving to deviate from classical foam ageing pathways and control the resulting morphologies.

\begin{acknowledgments}

The authors acknowledge F. Boulogne and S. Cabaret for the design of the quasi-2D cell.
This work was supported by the European Space Agency (ESA)
through project Soft Matter Dynamics and the French space
agency CNES with the project ‘‘Hydrodynamics of wet foams’’. A. Plati acknowledges funding from the Agence Nationale
de la Recherche (ANR), France, grant ANR-21-CE06-0039.
\end{acknowledgments}

\subsection*{End Matter}

\begin{figure}
\centering

\includegraphics[width=0.98\columnwidth,clip=true]{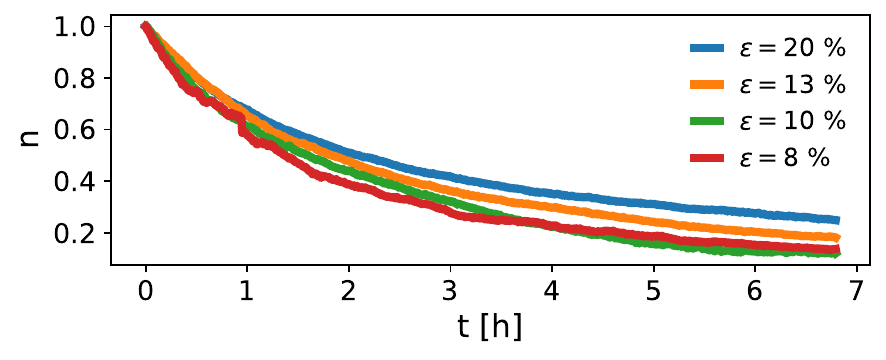}
\caption{Rescaled number of bubbles as a function of time for different liquid fractions in aqueous foams ($\phi=0$ $\%$) vibrated at amplitude $A=0.26$ mm.}
\label{fig:EndMatAcq}
\end{figure}

\emph{Vibrated aqueous foams---} To complement our investigation, we also performed some experiments on vibrated aqueous foam (\textit{i.e.} $\phi=0$ $\%$). Fig. \ref{fig:EndMatAcq} shows the behaviour of $n$ as a function of time for different liquid fractions under vibration at amplitude $A=0.26$ mm. An evident inflection point is never observed in these curves, signaling an ageing dynamics mainly dominated by coarsening. In all these experiments, coalescence occurs through rare events homogeneously distributed over time rather than through a cascade. This behaviour is qualitatively similar to that of a foamed emulsion with no yield stress (see the discussion of Figs.~\ref{fig:Fig1}c and d in the main text). Our experiments with aqueous foams provide further evidence
that triggering the coalescence catastrophe requires a non-zero continuous phase yield stress.

\begin{figure}
\centering
\includegraphics[width=0.98\columnwidth,clip=true]{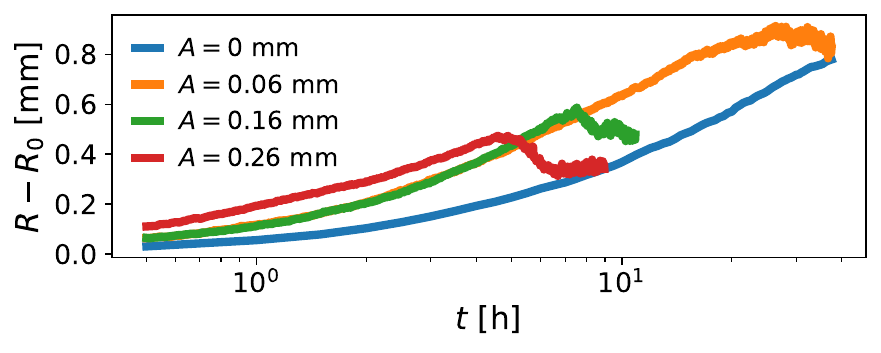}
\caption{Time evolution of the average bubble radius for the same experiments shown in Fig.~\ref{fig:Fig2}. $R_0$ defines the average bubble radius at $t_0$.}
\label{fig:EndMatterRad}
\end{figure}

\emph{Average bubble radius---}
In Fig.~\ref{fig:EndMatterRad} we plot the time evolution of the average bubble radius for the same experiments discussed in Fig.~\ref{fig:Fig2}. As we expect, this is a monotonously increasing function of time when coarsening is the dominant ageing mechanism. However, for $A=0.16$~mm and $A=0.26$~mm, it starts to decrease once the number of large bubbles arising from multiple coalescence events becomes sufficiently high. For $A=0.06$ mm, we observe a bump in the evolution of the average radius without any inflection in $n(t)$ (see Fig.~\ref{fig:Fig2}). This is due to coarsening being the dominating ageing mechanism (similar to the vibrated foamed emulsion at $\phi=65$ $\%$ and $A=0.26$ mm shown in Figs.~\ref{fig:Fig1}c and d).

\emph{Fitting procedure---}
Starting from the experimental curves $n(t)$, the parameters $a$, $b$ and $\sigma$ appearing in Eqs. \eqref{eq::modII} and \eqref{eq::pn} were obtained using the following procedure. First, we fitted $n(t)$ to the function $n(t)=(1+bt)^{-1}$ over the first hour of decay, during which coarsening dominates. This functional form corresponds to the solution of Eq. \eqref{eq::modII} with $a=0$, imposing $n(0)=1$. Next, we fitted the full decay curve, $n(t)$, with a 12$^{\rm{th}}$-order polynomial, whose analytical derivative provided us with $\dot{n}(t)$, and in turn with the curves $\dot{n}(n)$. These were fitted \textit{via} Eq. \eqref{eq::modII} to estimate $a$ and $\sigma$ using the previously obtained values of $b$. In some cases (see, for example, the green curve in Fig. \ref{fig:Fig2}), the time evolution of the rescaled number of bubbles stabilises at a non-zero value, $n_\text{f}$. This is to be expected given that the watershed algorithm can still identify some bubbles in the final foam structure. These bubbles do not evolve due to the large amount of residual continuous phase accumulated between them (see, for example, Fig. \ref{fig:Fig3}c). In such cases, we performed the fit after correcting Eq. \eqref{eq::modII} by shifting $n\to n-n_\text{f}$. This makes our model consistent with the fact that $\dot{n}(n_\text{f})=0$.

\bibliographystyle{apsrev4-1}
\bibliography{biblio}

\end{document}